\documentclass{article}
\usepackage[preprint,nonatbib]{nips_2018_mod}
\usepackage{amsmath}
\usepackage{amsthm}
\usepackage{array}
\usepackage{booktabs}
\usepackage{graphicx}
\usepackage{wasysym}
\usepackage{tabularx}
\usepackage{threeparttable}
\usepackage[newfloat,frozencache]{minted}  
\usepackage{caption}
\usepackage{subcaption}
\usepackage{parcolumns}
\usepackage{booktabs}
\usepackage{enumerate}

\usepackage{hyperref}
\hypersetup{pagebackref=false,breaklinks=true,colorlinks,bookmarks=false,urlcolor=blue,linkcolor=blue,citecolor=blue,pdftitle={ensmallen: a flexible C++ library for efficient function optimization},pdfauthor={Shikhar Bhardwaj, Ryan R. Curtin, Marcus Edel, Yannis Mentekidis, Conrad Sanderson}}

\usepackage[utf8]{inputenc} 
\usepackage[T1]{fontenc}    
\usepackage{booktabs}       
\usepackage{amsfonts}       
\usepackage{nicefrac}       
\usepackage{microtype}      

\usepackage{adjustbox}

\newcolumntype{R}[2]{%
  >{\adjustbox{angle=#1,lap=\width-(#2)}\bgroup}%
  l%
  <{\egroup}%
}
\newcommand*\rot{\multicolumn{1}{R{30}{2.0em}}}

\DeclareMathOperator*{\argmindown}{argmin}   
\DeclareMathOperator{\argminright}{argmin}   

\def\tinywhitedot{\textcolor{white}{\rm\tiny{.}}}

\makeatletter
\def\blfootnote{\xdef\@thefnmark{}\@footnotetext}
\makeatother

\begin{document}

\pagestyle{empty}

\title{\texttt{ensmallen}: a flexible C++ library for efficient function optimization}

\author{Shikhar Bhardwaj \\
Delhi Technological University \\
Delhi, India 110042 \\
\texttt{shikhar\_bt2k15@dtu.ac.in}
\And
Ryan R. Curtin \\
RelationalAI \\
Atlanta, GA, USA 30318 \\
\texttt{ryan@ratml.org}
\And
Marcus Edel \\
Free University of Berlin \\
Arnimallee 7, 14195 Berlin \\
\texttt{marcus.edel@fu-berlin.de}
\And
Yannis Mentekidis \\
Independent Researcher \\
\texttt{mentekid@gmail.com}
\And
Conrad Sanderson \\
Data61, CSIRO, Australia \\
University of Queensland, Australia\\
\texttt{conradsand{\tinywhitedot}@{\tinywhitedot}ieee.org}
}

\maketitle

\vspace*{-4ex}

\begin{abstract}
\vspace*{-1ex}
We present \texttt{\small ensmallen}, a fast and flexible C++ library for mathematical optimization of 
arbitrary user-supplied functions,
which can be applied to many machine learning problems.
Several types of optimizations are supported, including differentiable,
separable, constrained, and categorical objective functions.
The library provides many pre-built optimizers
(including numerous variants of SGD and Quasi-Newton optimizers)
as well as a flexible framework for implementing new optimizers and objective functions.
Implementation of a new optimizer requires only one method
and a new objective function requires typically one or two C++ functions. 
This can aid in the quick implementation and prototyping of new machine learning algorithms.
Due to the use of C++ template metaprogramming, \texttt{\small ensmallen} is able to
support compiler optimizations that provide fast runtimes.
Empirical comparisons show that \texttt{\small ensmallen} is able to outperform other
optimization frameworks (like Julia and SciPy), sometimes by large margins.
The library is distributed under the
BSD license and is ready for use
in production environments.
\end{abstract}

\blfootnote{\textbf{$^{\ast}$ Published in:} Workshop on Systems for ML and Open Source Software at NIPS / NeurIPS, 2018.}

\vspace*{-2ex}
\section{Introduction}
\vspace*{-1ex}

Mathematical optimization is the workhorse of virtually all machine learning
algorithms.  For a given objective function $f(\cdot)$
(which may have a special structure or constraints),
almost all machine learning problems can be boiled down
to the following optimization form:
\begin{equation}
\argmindown_x f(x).
\end{equation}
\vspace*{-1.5em}

Optimization is often computationally intensive and may correspond
to most of the time taken to train a machine learning model.  For instance, the
training of deep neural networks is dominated by the optimization of the model
parameters on the data~\cite{schmidhuber2015deep}.
Even popular machine learning models such as logistic regression
have training times mostly dominated by an optimization procedure~\cite{kingma2015adam}.

The ubiquity of optimization in machine learning algorithms highlights the need
for robust and flexible implementations of optimization algorithms.
We present \texttt{\small ensmallen}, a C++ optimization toolkit
that contains a wide variety of optimization techniques for many types of
objective functions.  Through the use of C++ template
metaprogramming~\cite{Abrahams_2004,Vandevoorde_2018},
\texttt{\small ensmallen} is able to generate efficient code that can help with the
demanding computational needs of many machine learning algorithms.

Although there are many existing machine learning optimization toolkits, few
are able to take explicit advantage of metaprogramming based code optimizations,
and few offer robust support for various types of objective functions.
For instance, deep learning
libraries like Caffe~\cite{jia2014caffe},
PyTorch~\cite{paszke2017automatic},
and TensorFlow~\cite{abadi2016tensorflow}
each contain a variety of optimization techniques.  However, these techniques are
limited to stochastic gradient descent (SGD) and SGD-like optimizers that
operate on small batches of data points at a time.  Other machine learning
libraries, such as \texttt{\small scikit-learn}~\cite{pedregosa2011scikit}
contain optimization algorithms but not in a coherent or reusable framework.
Many programming languages have higher-level packages for
mathematical optimization.  For example, \texttt{\small
scipy.optimize}~\cite{jones2014scipy},
is widely used in the Python community, and MATLAB's function optimization
support has been available and used for many decades.
However, these
implementations are often unsuitable for modern machine learning tasks---for
instance, computing the full gradient of the objective function may not be
feasible because there are too many data points.

In this paper, we describe the functionality of \texttt{\small ensmallen} and the types of
problems that it can be applied to.  We discuss the mechanisms by which \texttt{\small ensmallen} is
able to provide both computational efficiency and ease-of-use.
We show a few examples that use the library, as well as empirical performance comparisons
with other optimization libraries.

\texttt{\small ensmallen} is open-source software licensed under the 3-clause BSD
license~\cite{Rosen_2004_full}, 
allowing unencumbered use in both open-source and proprietary projects.
It is available for download from \url{https://ensmallen.org}.
Armadillo~\cite{sanderson2016armadillo} is used for efficient linear algebra operations,
with optional interface to GPUs via NVBLAS~\cite{nvidia2015}.

\vspace*{-0.2em}
\section{Types of Objective Functions}
\vspace*{-0.6em}

\begin{table}[b!]
\vspace*{-1.0em}
\centering
\begin{adjustbox}{scale={0.95}{0.95}}
    \begin{tabular}{@{} cl*{7}c @{}}
        & & \multicolumn{7}{c}{} \\[0.6ex]
        & & \rot{unified framework}
          & \rot{constraints}
          & \rot{batches}
          & \rot{arbitrary functions}
          & \rot{arbitrary optimizers}
          & \rot{sparse gradients}
          & \rot{categorical} \\
        \cmidrule[1pt]{2-9}
        & \texttt{\small ensmallen}            & \CIRCLE & \CIRCLE & \CIRCLE & \CIRCLE & \CIRCLE & \CIRCLE & \CIRCLE \\
        & Shogun \cite{sonnenburg2010shogun}             & \CIRCLE & - & \CIRCLE & \CIRCLE & \CIRCLE & - & -
\\
        & Vowpal Wabbit \cite{Langford2007VW}      & - & - & \CIRCLE  & - & - & - &
\CIRCLE \\
        & TensorFlow \cite{abadi2016tensorflow}        & \CIRCLE & -  & \CIRCLE  & \LEFTcircle & - &
\LEFTcircle & -  \\
        & Caffe \cite{jia2014caffe}           & \CIRCLE & -  & \CIRCLE & \LEFTcircle & \LEFTcircle
& - & - \\
        & Keras \cite{chollet2015}            & \CIRCLE & -  & \CIRCLE & \LEFTcircle & \LEFTcircle
& - & - \\
        & scikit-learn \cite{pedregosa2011scikit}       & \LEFTcircle & - & \LEFTcircle  & \LEFTcircle & -
& - & - \\
        & SciPy \cite{jones2014scipy}             & \CIRCLE & \CIRCLE  & -  & \CIRCLE & - & - & - \\
        & MATLAB \cite{mathworks2017OTB}            & \CIRCLE & \CIRCLE & - & \CIRCLE & - & - & - \\
        & Julia (\texttt{\small Optim.jl}) \cite{mogensen2018optim}         & \CIRCLE & - & - & \CIRCLE & - & - & - \\
        \cmidrule[1pt]{2-9}
    \end{tabular}
\end{adjustbox}
\caption{\footnotesize{
Feature comparison: \CIRCLE = provides feature,
\LEFTcircle = partially provides feature, - = does not provide feature.
{\it unified framework} indicates if there is some kind of generic/unified
optimization framework; {\it constraints} and {\it batches} indicate support for
constrained problems and batches; {\it arbitrary functions} means arbitrary
objective functions are easily implemented; {\it arbitrary optimizers} means
arbitrary optimizers are easily implemented; {\it sparse gradient} indicates
that the framework can natively take advantage of sparse gradients; and
{\it categorical} refers to if support for categorical features exists.
}}
\label{tab:functionality}
\vspace*{-1.8em}
\end{table}

\texttt{\small ensmallen} provides a {\bf set of optimizers} for
optimizing {\bf user-defined objective functions}.  It is also easy to implement a
new optimizer in the \texttt{\small ensmallen} framework.  Overall, our goal is to provide
an easy-to-use library that can solve the problem
$\argminright_{x} f(x)$
for any function $f(x)$ that takes a vector or matrix input $x$.
In most cases, $f(x)$ will have special structure; one example might be that
$f(x)$ is differentiable.  Therefore, the abstraction we have designed for \texttt{\small
ensmallen} can optionally take advantage of this structure.  For example, in
addition to $f(x)$, a user can provide an implementation of $f'(x)$, which in
turn allows first-order gradient-based optimizers to be used.  This generally
leads to significant speedups.

There are a number of other properties that \texttt{\small ensmallen} can use to
accelerate computation.  These are listed below:

\vspace*{-0.3em}
\begin{itemize} \itemsep -1pt
  \item {\bf arbitrary}: no assumptions can be made on $f(x)$
  \item {\bf differentiable}: $f(x)$ has a computable gradient $f'(x)$
  \item {\bf separable}: $f(x)$ is a sum of individual components: $f(x) =
\sum_{i} f_i(x)$
  \item {\bf categorical}: $x$ contains elements that can only take discrete
values
  \item {\bf sparse}: the gradient $f'(x)$ or $f'_i(x)$ (for a separable
function) is sparse
  \item {\bf partially differentiable}: the separable gradient $f_i'(x)$ is also
separable for a different axis $j$
  \item {\bf bounded}: $x$ is limited in the values that it can take
\end{itemize}
\vspace*{-0.3em}

Due to its straightforward abstraction framework, \texttt{\small ensmallen} is able to
provide a large set of diverse optimization algorithms for objective functions
with these properties.  Below is a list of currently available optimizers:

\vspace*{-0.4em}
\begin{enumerate}[{~~~$\bullet$}]
\small
  \item {\bf SGD variants}~\cite{Ruder_2016}:  
      Stochastic Gradient Descent (SGD),
      SGD with Restarts,
      Parallel SGD (Hogwild!)~\cite{hogwild_nips_2011},
      Stochastic Coordinate Descent, 
      AdaGrad~\cite{Duchi_2011},
      AdaDelta~\cite{Zeiler_2012},
      RMSProp,
      SMORMS3,
      Adam~\cite{kingma2015adam},
      AdaMax

  \item {\bf Quasi-Newton variants}:
      Limited-memory BFGS (L-BFGS)~\cite{zhu1997algorithm},
      incremental Quasi-Newton method~\cite{Mokhtari_2018},
      Augmented Lagrangian Method~\cite{Hestenes_1969}

  \item {\bf Genetic variants}:
      Conventional Neuro-evolution~\cite{Belew_1990},
      Covariance Matrix Adaptation Evolution Strategy~\cite{Hansen_2001}

  \item {\bf Other}:
    Conditional Gradient Descent,
    Frank-Wolfe algorithm~\cite{Frank_1956},
    Simulated Annealing~\cite{kirkpatrick1983optimization}
\end{enumerate}

In \texttt{\small ensmallen}'s framework, if a user wants to optimize a differentiable objective
function, they only need to provide implementations of $f(x)$ and $f'(x)$, and
then they can use any of the gradient-based optimizers that \texttt{\small ensmallen}
provides.  Table~\ref{tab:functionality} contrasts 
the classes of objective functions that can be handled by \texttt{\small ensmallen}
and other popular frameworks and libraries.

Not every optimization algorithm provided by \texttt{\small ensmallen} can be
used by every class of objective function; for instance, a gradient-based
optimizer such as L-BFGS cannot operate on a non-differentiable objective
function.
Thus, the best the library can attain is to maximize the flexibility
available, so that a user can easily implement a function $f(x)$ and have it
work with as many optimizers as possible.

To accomplish the flexibility, \texttt{\small ensmallen} makes heavy use of
C++ template metaprogramming.
When implementing an objective function to be optimized,
a user may only need to implement a few methods; metaprogramming is then automatically
used to check that the given functions match the requirements of the
optimizer that is being used.  When implementing an optimizer, we can assume
that the given function to be optimized meets the required assumptions of the
optimizers, and encode those requirements as compile-time checks
(via \texttt{\small static\_assert}); this can provide much easier-to-understand
error messages than typical compiler output for templated C++ code.

For the most common case of a differentiable $f(x)$, the user only needs to
implement two methods:

\vspace*{-0.3em}
\begin{itemize} \itemsep -1pt
  \item \texttt{\small double Evaluate($x$)}: given coordinates $x$, this function
returns the value of $f(x)$.
  \item \texttt{\small void Gradient($x$, $g$)}: given coordinates $x$ and a reference
to $g$, set $g = f'(x)$.
\end{itemize}
\vspace*{-0.3em}

Alternatively, the user can simply implement a \texttt{\small EvaluateWithGradient()}
function that computes both $f(x)$ and $f'(x)$ simultaneously, which is useful
in cases where both the objective and gradient depend on similar computations.

The required API for separable differentiable objective functions (i.e.~those
that would use an optimizer like SGD) is very similar, except that
\texttt{\small Evaluate()}, \texttt{\small Gradient()} and \texttt{\small EvaluateWithGradient()}
should operate only on mini-batches, and utility methods \texttt{\small Shuffle()} and
\texttt{\small NumFunctions()} must be added.  The same pattern applies for other
types of objective functions: only a few methods specific to class of objective
function itself must be implemented and then any optimizer may be used.

\vspace*{-0.3em}
\section{Example: Learning Linear Regression Models}
\label{sec:linreg_example}
\vspace*{-0.5em}

As an example of usage, consider the linear regression objective
function\footnote{For simplicity, we ignore the bias term.  It can be
rederived by taking $x^*_i = (x_i, 1)$.}.  Given a dataset $X \in
\mathcal{R}^{n \times d}$ and associated responses $y \in \mathcal{R}^n$, the
model of linear regression is to assume that $y_i = x_i \theta$ for each
point and response $(x_i, y_i)$.  To fit this model $\theta \in \mathcal{R}^d$
to the data, we must find

\vspace*{-0.5em}
\begin{equation}
\argmindown_\theta f(\theta) =
\argmindown_\theta \sum\nolimits_{i = 1}^n (y_i - x_i \theta)^2 =
\argmindown_\theta \| y - X \theta \|_F^2.
\end{equation}
\vspace*{-0.5em}

The objective function $f(\theta)$ has the associated gradient

\vspace*{-0.5em}
\begin{equation}
f'(\theta) = \sum\nolimits_{i = 1}^n -2 x_i (y_i - x_i \theta) = -2 X^T (y - X \theta).
\end{equation}
\vspace*{-0.5em}

We can implement these two functions in a class named \texttt{\small LinearRegressionFunction},
as shown in Fig.~\ref{fig:LinearRegressionFunction}.
This is the entire required implementation to optimize the linear regression model with
any of the gradient-based optimizers in \texttt{\small ensmallen}.

\begin{figure}[!tb]
\hrule\vspace*{0.5ex}
\begin{adjustbox}{scale={0.95}{0.90}}
\begin{minipage}{\textwidth}
\begin{minted}[fontsize=\small]{c++}
class LinearRegressionFunction {
 public:
  // Construct the LinearRegressionFunction with the given data
  LinearRegressionFunction(arma::mat& X_in, arma::vec& y_in) : X(X_in), y(y_in) {}

  // Compute the objective function
  double Evaluate(const arma::mat& theta) {
    return std::pow(arma::norm(y - X * theta), 2.0);
  }
  // Compute the gradient and store in 'gradient'
  void Gradient(const arma::mat& theta, arma::mat& gradient) {
    gradient = -2 * X.t() * (y - X * theta);
  }
 private:
  arma::mat& X; arma::vec& y;
};
\end{minted}
\end{minipage}
\end{adjustbox}
\vspace*{0.5ex}\hrule\vspace*{0.5ex}
\caption
  {
  Implementation of objective and gradient functions for linear regression,
  used by optimizers in \texttt{ensmallen}.
  The types {\footnotesize\tt arma::mat} and {\footnotesize\tt arma::vec}
  are matrix and vector types
  from Armadillo~\cite{sanderson2016armadillo}.
  }
\label{fig:LinearRegressionFunction}
\end{figure}

Given the user-defined \texttt{\small LinearRegressionFunction} class,
the code snippet below
shows how the L-BFGS optimizer can be used to find the best parameters $\theta$:

\vspace*{-0.4em}
\begin{minted}[fontsize=\small]{c++}
    LinearRegressionFunction lrf(X, y); // we assume X and y already hold data
    ens::L_BFGS lbfgs; // create L-BFGS optimizer with default parameters

    arma::vec theta(X.n_rows, arma::fill::randu); // random uniform initialization
    lbfgs.Optimize(lrf, theta); // after this call, theta holds the solution
\end{minted}
\vspace*{-0.4em}

To use the small-batch SGD-like optimizers provided by \texttt{\small ensmallen},
only a slight variation on the signature of \texttt{\small Evaluate()} and
\texttt{\small Gradient()} would be needed, plus the \texttt{\small Shuffle()}
and \texttt{\small NumFunctions()} utility methods.

\vspace*{-0.3em}
\section{Automatic Metaprogramming for Ease of Use and Efficiency}
\vspace*{-0.5em}

When optimizing a given function $f(x)$, the computation of
the objective function $f(x)$ and its derivative $f'(x)$ often involve the
computation of identical intermediate results.  Consider the linear regression
objective function described
in Section~\ref{sec:linreg_example}, {\small $f(\theta) = \| y - X\theta \|_F^2$}.
For this objective function, the derivative $f'(\theta)$ has a related form of
{\small $-2 X^T (y -X \theta)$}.  Both the objective function and the derivative 
depend on the computation of the vector term {\small $(y - X \theta)$},
which can be computationally expensive to compute if $X$ is a large matrix.
Existing optimization frameworks do not have an easy way to avoid
this duplicate computation. In many cases, an optimization algorithm
may need the values of both $f(\theta)$ and $f'(\theta)$ for a given $\theta$.

Using template metaprogramming, \texttt{\small ensmallen} provides an easy (and
optional) way for users to avoid this extra computational overhead.  Instead of
specifying individual \texttt{\small Evaluate()} and \texttt{\small Gradient()} functions, a user
may simply write an \texttt{\small EvaluateWithGradient()} function that returns both the
objective value and the gradient value for an input $\theta$.  As an example,
for the 
\texttt{LinearRegressionFunction} class in Fig.~\ref{fig:LinearRegressionFunction},
we can replace \texttt{\small Evaluate()} and \texttt{\small Gradient()}
with an implementation of \texttt{\small EvaluateWithGradient()}
that computes {\small $(y - X \theta)$} only once:

\begin{adjustbox}{scale={0.95}{0.95}}
\begin{minipage}{1\textwidth}
\begin{minted}[fontsize=\small]{c++}
    double EvaluateWithGradient(const arma::mat& theta, arma::mat& gradient) {
      const arma::vec v = (y - X * theta); // Cache result
      gradient = -2 * X.t() * v;  // Store gradient in the provided matrix
      return arma::accu(v % v); // Take squared norm of v
    }
\end{minted}
\end{minipage}
\end{adjustbox}

Template metaprogramming techniques are automatically used to
detect which methods exist, and a wrapper class will use suitable mix-ins in
order to provide `missing` functionality~\cite{smaragdakis2000mixin}.  For
instance, if \texttt{\small EvaluateWithGradient()} is not provided, a version will be
automatically generated that calls both \texttt{\small Gradient()} and \texttt{\small Evaluate()} in turn.
Similarly, if \texttt{\small Evaluate()} or \texttt{\small Gradient()} does not exist, then \texttt{\small
EvaluateWithGradient()} is called, and the unnecessary part of the result will
be discarded.

The use of template metaprogramming in this manner also allows for compiler optimizations that
would not otherwise be possible (and that are often not possible in other
frameworks).  Firstly, because the objective function class itself is a template
parameter, the compiler is able to avoid the overhead of a function pointer
dereference, which would not be easily possible when using a language with
virtual inheritance.  The compiler is also able to use inlining and any
optimizations that may imply, including removing temporary values and dead code
elimination.  Further, if \texttt{\small ensmallen} automatically generates an
\texttt{\small Evaluate()} or \texttt{\small Gradient()} method from a user-supplied
\texttt{\small EvaluateWithGradient()} method, the compiler can in some cases recognize and
remove the computation of unnecessary results.  For instance, in an
automatically generated \texttt{\small Evaluate()} method, the computation of the gradient
from \texttt{\small EvaluateWithGradient()} can be avoided entirely.

Overall, the automatic code generation functionality in \texttt{\small ensmallen}
reduces the requirements for users
when they are implementing their own objective functions to be optimized,
and allows users a way to provide more efficient implementations of their
objective functions.
This leads to quicker development, quicker results, and reduces the likelihood of bugs.

At the time of writing, the automatic code generation
is implemented for the most commonly-used cases:
full-batch and small-batch \texttt{\small Evaluate()}, \texttt{\small Gradient()},
and \texttt{\small EvaluateWithGradient()}.  We aim to expand this support to other
sets of methods for other types of objective functions.

\vspace*{-0.3em}
\section{Experiments}
\vspace*{-0.5em}

\begin{table}[t]
\begin{center}
\begin{tabular}{lcccc}
\toprule
 & \texttt{\small ensmallen} & \texttt{\small scipy} & \texttt{\small Optim.jl} & \texttt{\small samin} \\
\midrule
default & {\bf 0.004s} & 1.069s & 0.021s & 3.173s \\
tuned & & 0.574s & & 3.122s \\
\bottomrule
\end{tabular}
\end{center}
\caption{Runtimes for $100000$ iterations of simulated annealing with various
frameworks on the simple Rosenbrock function.  Julia code runs do not count
compilation time.  The {\it tuned} row indicates that the code was manually
modified for speed.}
\label{tab:rosenbrock_results}
\end{table}

To demonstrate the benefits of the metaprogramming based code optimizations
that \texttt{\small ensmallen} can exploit,
we compare the performance of \texttt{\small ensmallen} with several other
optimization frameworks, including some that use automatic differentiation.

For our first experiment, we consider the simple and popular Rosenbrock
function~\cite{Rosenbrock1960}: $f([x_1, x_2]) = 100 (x_2 - x_1^2)^2 + (1 -
x_1^2)$.  For the optimizer, we use simulated
annealing~\cite{kirkpatrick1983optimization}, a gradient-free optimizer.
Simulated annealing will call the objective function numerous times; for each
simulation we limit the optimizer to 100k objective evaluations.  Since the
objective function is straightforward and is called many times, this can help us
understand the overheads of various frameworks.

We compare four frameworks for this task:
{\bf (i)}~\texttt{\small ensmallen},
{\bf (ii)}~\texttt{\small scipy.optimize.anneal} from SciPy 0.14.1~\cite{jones2014scipy},
{\bf (iii)}~simulated annealing implementation in \texttt{\small Optim.jl} with Julia 1.0.1~\cite{mogensen2018optim},
{\bf (iv)}~\texttt{\small samin} in the \texttt{\small optim} package for Octave~\cite{octave}.
While another option here might be \texttt{\small simulannealbnd()} 
in the Global Optimization Toolkit for MATLAB,
no license was available.
We ran our code on a MacBook Pro i7 2018 with 16GB RAM running macOS 10.14 with clang 1000.10.44.2, Julia version 1.0.1, Python 2.7.15, and Octave 4.4.1.

Initially, we implemented these functions as simply as possible and ran them
without any tuning. This reflects how a typical user might interact with a
given framework.
The results are shown in the first row of
Table~\ref{tab:rosenbrock_results}.  Only Julia and \texttt{\small
ensmallen} are compiled, and thus are able to avoid the function pointer
dereference and take advantage of inlining and related optimizations.

However, both Python and Octave have routes for acceleration,
such as Numba~\cite{lam2015numba}, MEX bindings and JIT compilation.
We hand-optimized the Rosenbrock implementation using Numba,
which required significant modification of the
underlying \texttt{\small anneal.anneal()} function.
These techniques did produce some speedup,
as shown in the second row of Table~\ref{tab:rosenbrock_results}.
For Octave, a MEX binding did not produce a noticeable difference.
We were unable to tune either \texttt{\small ensmallen} or \texttt{\small
Optim.jl} to get any speedup, suggesting that novice users will easily be able
to write efficient code in these cases.

Next, we consider the linear regression example described in
Sec.~\ref{sec:linreg_example}.  For this task we use the first-order L-BFGS
optimizer~\cite{zhu1997algorithm}.  Using the same four packages, we implement
the linear regression objective and gradient.  For \texttt{\small ensmallen} we
implement a version with only \texttt{\small EvaluateWithGradient()},
denoted as \texttt{\small ensmallen-1}.  We also implement a version with both
\texttt{\small Evaluate()} and \texttt{\small Gradient()}: \texttt{\small ensmallen-2}.
We also use automatic differentiation for Julia via the \texttt{\small
ForwardDiff.jl}~\cite{RevelsLubinPapamarkou2016} package and for Python via the
\texttt{\small Autograd}~\cite{maclaurin2015autograd}
package.  For GNU Octave we use the \texttt{\small bfgsmin()} function.

Results for various data sizes are shown in Table~\ref{tab:lbfgs}.  For each
implementation, L-BFGS was allowed to run for only $10$ iterations and never
converged in fewer iterations.  The datasets used for training are highly noisy random
data with a slight linear pattern. Note that the exact data is not relevant
for the experiments here, only its size.  Runtimes are reported as the
average across 10 runs.

\begin{table}
\centering
\begin{adjustbox}{scale={0.90}{0.90}}
\begin{tabular}{lccccc}
\toprule
{\em algorithm} & $d$: 100, $n$: 1k & $d$: 100, $n$: 10k & $d$: 100, $n$:
100k & $d$: 1k, $n$: 100k \\
\midrule
\texttt{\small ensmallen}-1 & {\bf 0.001s} & {\bf 0.009s} & {\bf 0.154s} & {\bf 2.215s} \\
\texttt{\small ensmallen}-2 & 0.002s & 0.016s & 0.182s & 2.522s \\
\texttt{\small Optim.jl} & 0.006s & 0.030s & 0.337s & 4.271s \\
\texttt{\small scipy} & 0.003s & 0.017s & 0.202s & 2.729s \\
\texttt{\small bfgsmin} & 0.071s & 0.859s & 23.220s & 2859.81s\\
\texttt{\small ForwardDiff.jl} & 0.497s & 1.159s & 4.996s & 603.106s \\
\texttt{\small autograd} & 0.007s & 0.026s & 0.210s & 2.673s \\
\bottomrule
\end{tabular}
\end{adjustbox}
\vspace*{0.25ex}
\caption{\footnotesize
Runtimes for the linear regression function on various dataset sizes,
with $n$ indicating the number of samples,
and $d$ indicating the dimensionality of each sample.
All Julia runs do not count compilation time.}
\label{tab:lbfgs}
\end{table}

The results indicate that \texttt{\small ensmallen} with \texttt{\small
EvaluateWithGradient()} is the fastest approach.
Furthermore, the use of \texttt{\small EvaluateWithGradient()} yields considerable
speedup over the \texttt{\small ensmallen-2} implementation with both the
objective and gradient implemented separately.  In addition, although the
automatic differentiation support makes it easier for users to write their
code, we observe that the output of automatic differentiators is not as
efficient, especially with \texttt{\small ForwardDiff.jl}.  We expect this
effect to be more pronounced with increasingly complex objective functions.

\begin{figure}[b!]
\centering
\begin{subfigure}[b]{0.49\textwidth}
\begin{minted}[fontsize=\footnotesize]{c++}
// X and y are data
LinearRegressionFunction lrf(X, y);

using namespace ens;
StandardSGD<>().Optimize(lrf, sgdModel);
Adam().Optimize(lrf, adamModel);
AdaGrad().Optimize(lrf, adagradModel);
SMORMS3().Optimize(lrf, smorms3Model);
SPALeRASGD().Optimize(lrf, spaleraModel);
RMSProp().Optimize(lrf, rmspropModel);
\end{minted}
\caption{Code.}
\end{subfigure}
\begin{subfigure}[b]{0.49\textwidth}
  \centering
  \includegraphics[width=\textwidth,height=0.6\textwidth]{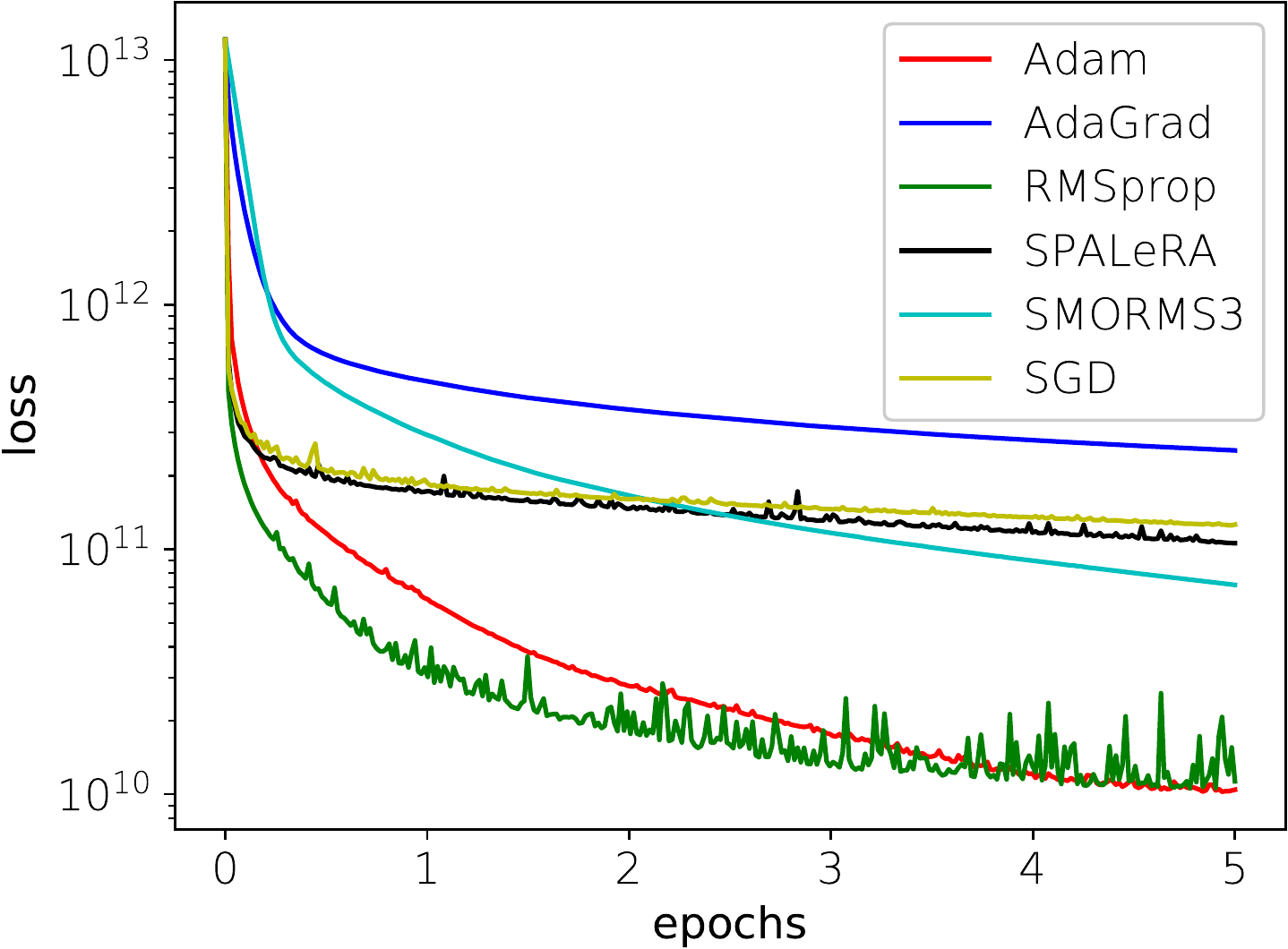}
\caption{Learning curves.}
\end{subfigure}
\caption{\footnotesize
Example usage of six \texttt{\small ensmallen} optimizers to optimize a
linear regression function on the \texttt{\small YearPredictionMSD}
dataset~\cite{ucimlrepository} for 5 epochs of training.  The optimizers can be
tuned for better performance.}
\label{fig:learning_curve}
\vspace*{-1ex}
\end{figure}

Lastly, we demonstrate the easy pluggability in \texttt{\small ensmallen}
for using various optimizers on the same task.
Using a version of \texttt{\small LinearRegressionFunction} from Sec.~\ref{sec:linreg_example}
adapted for separable differentiable optimizers,
we run six optimizers with default parameters in just 8 lines of code,
as shown in Fig.~\ref{fig:learning_curve}(a).
Applying these optimizers to the \texttt{\small YearPredictionMSD}
dataset from the UCI repository~\cite{ucimlrepository}
yields the learning curves shown in Fig.~\ref{fig:learning_curve}(b).
Any other optimizer for separable differentiable objective
functions can be dropped into place in the same manner.

\vspace*{-0.3em}
\section{Conclusion}
\vspace*{-0.5em}

We have described \texttt{\small ensmallen}, a flexible C++ library for function
optimization that provides an easy interface for the implementation and optimization
of user-defined objective functions.  Many types of functions can be optimized,
including separable and constrained functions.
The library provides many pre-built optimizers (including numerous variants
of SGD and Quasi-Newton optimizers).
The library internally exploits template metaprogramming
to maximise opportunities for efficiency gains,
as well as 
to make the implementation of objective functions easier
by automatically generating missing methods.

We aim to expand the library with further optimization techniques
as the need arises.  Since \texttt{\small ensmallen} is open source,
external contributions to the codebase are welcome.
For more information on the
library, see the website and documentation at
\href{https://ensmallen.org}{\texttt{\small https://ensmallen.org}}.
The library is already in use for function optimization in the
\texttt{\small mlpack} machine learning toolkit~\cite{mlpack2018}.

{\bf Acknowledgements.}
We would like to thank the many contributors to \texttt{\small ensmallen},
who are listed on the associated website.

\bibliographystyle{abbrv}
\bibliography{refs}

\end{document}